\documentclass[a4paper,11pt]{article}
\usepackage{pos}

\usepackage{textcomp}
\usepackage{amsfonts} % AMS
\usepackage{amsmath} % AMS
\usepackage{slashed}
\usepackage{pifont}
\usepackage{multirow,lscape}
\usepackage{array}
\usepackage{subcaption}
\usepackage{verbatim}
\usepackage{bm}
\usepackage{comment}
\usepackage{xcolor}
\usepackage{url} %%% URL
\graphicspath{{./fig/}}

\providecommand{\wbar}[1]{\overline#1}

\providecommand{\mate}[3]{\langle#1\lvert#2\rvert#3\rangle}

\renewcommand{\Re}{\mathrm{Re}\,}
\renewcommand{\Im}{\mathrm{Im}\,}

\providecommand{\GeV}{\;\mathrm{GeV}}

\definecolor{HLBlue}{HTML}{6599FF}
\definecolor{HLOrange}{HTML}{FF6600}

\newcommand{\BK}{\hat{B}_{K}}
\newcommand{\Vcb}{|V_{cb}|}

\newcommand{\Vus}{|V_{us}|}

\newcommand{\eps}{\varepsilon}

\newcommand{\epsK}{\varepsilon_{K}}

\newcommand{\BtoDstp}{\bar{B} \to D^{(\ast)} \ell \bar{\nu}}

\newcommand{\wlee}[1]{\textcolor{red}{#1}} % highlight wlee
\newcommand{\red}[1]{\textcolor{red}{#1}} 
\newcommand{\blue}[1]{\textcolor{blue}{#1}}

\begin{document}
\title{2021 Update on $\epsK$ with lattice QCD inputs}
\ShortTitle{$\epsK$ with lattice QCD inputs}

\author[a,1]{Jeehun Kim}
\author[b,1]{Yong-Chull Jang}
\author[a,1]{Sunkyu Lee}
\author*[a,1]{Weonjong Lee}
\author[c,1]{Jaehoon Leem}
\author[a,1]{Chanju Park}
\author[d,1]{Sungwoo Park}

\affiliation[a]{Lattice Gauge Theory Research Center, CTP, and FPRD,
  Department of Physics and Astronomy, \\
  Seoul National University,
  Seoul 08826, South Korea}

\affiliation[b]{Columbia University,
  Department of Physics,
  538 West 120th Street,
  New York, NY 10027, USA}

\affiliation[c]{School of Physics,
  Korea Institute for Advanced Study (KIAS),
  Seoul 02455, South Korea}

\affiliation[d]{
  Thomas Jefferson National Accelerator Facility,
  12000 Jefferson Avenue,
  Newport News, VA 23606, USA}

\emailAdd{wlee@snu.ac.kr}

\note{The SWME collaboration}

% obsolete options
%\keywords{flavor physics, CP violation, standard model phenomenology}
%\arxivnumber{1111.1111}

\abstract{ We present recent updates for $\epsK$ determined directly
  from the standard model (SM) with lattice QCD inputs such as $\BK$,
  $\Vcb$, $\Vus$, $\xi_0$, $\xi_2$, $\xi_\text{LD}$, $f_K$, and $m_c$.
  We find that the standard model with exclusive $\Vcb$ and other
  lattice QCD inputs describes only 66\% of the experimental value of
  $|\epsK|$ and does not explain its remaining 34\%, which leads to a
  strong tension in $|\epsK|$ at the $4.5\sigma \sim 3.7\sigma$ level
  between the SM theory and experiment.  We also find that this
  tension disappears when we use the inclusive value of $\Vcb$
  obtained using the heavy quark expansion based on the QCD sum rule
  approach. }

\FullConference{ The 38th International Symposium on Lattice Field Theory,
  LATTICE2021
  26th-30th July, 2021
  Zoom/Gather@Massachusetts Institute of Technology}

\maketitle

\section{Introduction}
This paper is an update of our previous papers \cite{ Kim:2019vic,
  Bailey:2018feb, Bailey:2015tba, Bailey:2018aks, Jang:2017ieg,
  Bailey:2015frw}.
Here, we present recent progress in determination of $|\epsK|$ with
updated inputs from lattice QCD.

Here, we follow the color convention of our previous papers \cite{
  Kim:2019vic, Bailey:2018feb, Bailey:2015tba, Bailey:2018aks,
  Jang:2017ieg, Bailey:2015frw} in Tables
\ref{tab:xi0-sum}--\ref{tab:epsK}.
We use the red color for the new input data which is used to evaluate
$\epsK$.
We use the blue color for the new input data which is not used for
some obvious reason.

\section{Input parameter $\xi_0$}
The absorptive part of long distance effects on $\epsK$ is parametrized
into $\xi_0$.
\begin{align}
  \xi_0  &= \frac{\Im A_0}{\Re A_0}, \qquad
  \xi_2 = \frac{\Im A_2}{\Re A_2}, \qquad
  \Re \left(\frac{\eps'}{\eps} \right) =
  \frac{\omega}{\sqrt{2} |\eps_K|} (\xi_2 - \xi_0) \,.
  \label{eq:e'/e:xi0}
\end{align}
There are two independent methods to determine $\xi_0$ in lattice QCD:
the indirect and direct methods.
The indirect method is to determine $\xi_0$ using
Eq.~\eqref{eq:e'/e:xi0} with lattice QCD results for $\xi_2$ combined
with experimental results for $\eps'/\eps$, $\epsK$, and $\omega$.
The direct method is to determine $\xi_0$ directly using the lattice
QCD results for $\Im A_0$, combined with experimental results for $\Re
A_0$.

In Table~\ref{tab:xi0-sum} (\subref{tab:exp-ReA0-ReA2-1}), we summarize
experimental results for $\Re A_0$ and $\Re A_2$.
In Table~\ref{tab:xi0-sum} (\subref{tab:ImA0-ImA2-1}), we summarize
lattice results for $\Im A_0$ and $\Im A_2$ calculated by RBC-UKQCD.
In Table~\ref{tab:xi0-sum} (\subref{tab:xi0-1}), we summarize results
for $\xi_0$ which is obtained using results in Table~\ref{tab:xi0-sum}
(\subref{tab:exp-ReA0-ReA2-1}) and (\subref{tab:ImA0-ImA2-1}).

Here, we use results of the indirect method for $\xi_0$ to evaluate
$\epsK$, since its systematic and statistical errors are much smaller
than those of the direct method.

\begin{table}[htbp]
  \begin{subtable}{1.0\linewidth}
    \renewcommand{\arraystretch}{1.2}
    \center
%    \vspace*{-7mm}
    \resizebox{1.0\textwidth}{!}{
      \begin{tabular}{ @{\qquad} l @{\qquad} l @{\qquad\qquad} l @{\qquad\qquad} c @{\qquad} l @{\qquad\qquad} }
        \hline\hline
        parameter & method & value & Ref. & source \\ \hline
        $\Re A_0$ & exp & \red{ $3.3201(18) \times 10^{-7} \GeV$ } &
        \cite{ Blum:2015ywa, Bai:2015nea}  & NA
        \\
        $\Re A_2$ & exp & \red{ $1.4787(31) \times 10^{-8} \GeV$ } &
        \cite{ Blum:2015ywa} & NA
        \\ \hline
        $\omega$ & exp & $0.04454(12)$ &
        \cite{ Blum:2015ywa} & NA
        \\ \hline
        $|\epsK|$ & exp & $2.228(11) \times 10^{-3}$ &
        \cite{ Zyla:2020zbs} & PDG-2021
        \\
        $\Re(\eps'/\eps)$ & exp & $1.66(23) \times 10^{-3}$ &
        \cite{ Zyla:2020zbs} & PDG-2021
        \\ \hline\hline
      \end{tabular}
    } %%% \resizebox
    \caption{Experimental results for $\omega$, $\Re A_0$ and $\Re A_2$.}
    \label{tab:exp-ReA0-ReA2-1}
  \end{subtable}
  \vspace*{3mm}
  \begin{subtable}{1.0\linewidth}
    \renewcommand{\arraystretch}{1.2}
    \center
%    \vspace*{-7mm}
    \resizebox{1.0\textwidth}{!}{
      \begin{tabular}{ @{\qquad} l @{\qquad} l @{\qquad}@{\qquad} l @{\qquad}@{\qquad} c @{\qquad} l }
        \hline\hline
        parameter & method & value ($\GeV$) & Ref. & source \\ \hline
        $\Im A_0$ & lattice & \red{$-6.98(62)(144) \times 10^{-11}$} &
        \cite{ RBC:2020kdj}  & RBC-UK-2020 p4t1
        \\
        $\Im A_2$ & lattice & \red{$-8.34(103) \times 10^{-13}$}  &
        \cite{ RBC:2020kdj} & RBC-UK-2020 p31e90
        \\ \hline\hline
      \end{tabular}
    } %%% \resizebox
    \caption{Results for $\Im A_0$, and $\Im A_2$ in lattice QCD. }
    \label{tab:ImA0-ImA2-1}
  \end{subtable}
  \vspace*{3mm}
  \begin{subtable}{1.0\linewidth}
    \renewcommand{\arraystretch}{1.2}
    \center
%    \vspace*{-7mm}
    \resizebox{1.0\textwidth}{!}{
      \begin{tabular}{@{\qquad} l @{\qquad\qquad} l @{\qquad\qquad} l @{\qquad\qquad} l @{\qquad\qquad} l @{\qquad} }
        \hline\hline
        parameter & method & value & ref & source \\ \hline
        $\xi_0$ & indirect & $-1.738(177) \times 10^{-4}$ & \cite{ RBC:2020kdj} & SWME \\
        $\xi_0$ & direct  & $-2.102(472) \times 10^{-4}$  & \cite{ RBC:2020kdj} & SWME \\ \hline\hline
      \end{tabular}
    } %%% \resizebox
    \caption{Results for $\xi_0$ obtained using the direct and indirect
      methods in lattice QCD. }
    \label{tab:xi0-1}
  \end{subtable}
  \caption{Results for $\xi_0$. The p4t1 is an abbreviation for Table 1 in page 4. The p31e90 is an abbreviation for Eq.~(90) in page 31.}
  \label{tab:xi0-sum}
\end{table}

\section{Input parameters: $\Vcb$}
\label{sec:Vcb}
In Table \ref{tab:Vcb} (\subref{tab:ex-Vcb}) and
(\subref{tab:in-Vcb}), we present recent updates for exclusive $\Vcb$
and inclusive $\Vcb$ respectively.
In Table \ref{tab:Vcb} (\subref{tab:ex-Vcb}), we summarize results for
exclusive $\Vcb$ obtained by various groups: HFLAV, BELLE, BABAR,
FNAL/MILC, LHCb, and FLAG.
Results from LHCb comes from analysis on $B_s\to D^*_s \ell \bar{\nu}$
decays which are not available in the $B$-factories.
Since the decays modes of $B_s$ have poor statistics, the final
results have overall uncertainty much larger than those of $B$ by an
order of magnitude.
Hence, we drop out results of LHCb in this article without loss of
fairness.
The rest of results for exclusive $\Vcb$ have comparable size of
errors and are consistent with one another within $1.0\sigma$
statistical uncertainty.
In addition, it is nice to observe all the results be consistent
between the CLN and BGL analysis, after all the boisterous debates
\cite{ Bailey:2018feb, FermilabLattice:2021cdg}.

In Table \ref{tab:Vcb} (\subref{tab:in-Vcb}), we present recent
results for inclusive $\Vcb$.
The HFLAV group has reported the same results for inclusive $\Vcb$
in 2021 as in 2017, while FLAG reported updated results.
\begin{table}[t!]
  \begin{subtable}{1.0\linewidth}
    \renewcommand{\arraystretch}{1.2}
    \center
    \vspace*{-5mm}
    \resizebox{1.0\textwidth}{!}{
      \begin{tabular}{@{\qquad} l @{\qquad} l @{\qquad} l @{\qquad} l @{\qquad} l @{\qquad}}
        \hline\hline
        channel & value & method & ref & source \\ \hline
%----------
% HFLAV
%----------
        ex-comb & $39.13(59)$ & comb & \cite{Amhis:2016xyh} & HFLAV-2017 
        \\
        ex-comb & \red{$39.25(56)$} & CLN & \cite{HFLAV:2019otj} p115e223 & HFLAV-2021  
        \\ \hline
%----------
% BELLE
%----------
        $B\to D^* \ell \bar{\nu}$
        & \red{$39.0(2)(6)(6)$} & CLN & \cite{Belle:2018ezy} erratum p4 & BELLE-2021 \\
        $B\to D^* \ell \bar{\nu}$
        & \red{$38.9(3)(7)(6)$} & BGL & \cite{Belle:2018ezy} erratum p4 & BELLE 2021
        \\ \hline
%----------
% BABAR
%----------
        $B\to D^* \ell \bar{\nu}$
        & \red{$38.40(84)$} & CLN & \cite{ BaBar:2019vpl} p5t2 & BABAR-2019
        \\
        $B\to D^* \ell \bar{\nu}$
        & \red{$38.36(90)$} & BGL & \cite{ BaBar:2019vpl} p5t1 & BABAR-2019
        \\ \hline
%---------------------------
% FNAL/MILC (BELLE + BABAR) 
%---------------------------        
        $B\to D^* \ell \bar{\nu}$
        & \red{$38.57(78)$} & BGL & \cite{ FermilabLattice:2021cdg} &
        FNAL/MILC-2021
        \\
        & & & p27e5.22, p34e6.1 &
        \\ \hline
%----------
% LHCb 
%----------
        $B_s\to D^*_s \ell \bar{\nu}$
        & \blue{$41.4(6)(9)(12)$} & CLN & \cite{ LHCb:2020cyw} p15 & LHCb-2020
        \\
        $B_s\to D^*_s \ell \bar{\nu}$
        & \blue{$42.3(8)(9)(12)$} & BGL & \cite{ LHCb:2020cyw} p15 & LHCb-2020
        \\ \hline
%----------
% FLAG 
%----------
        ex-comb
        & \textcolor{red}{$39.48(68)$} & comb & \cite{ Aoki:2021kgd} p191 & FLAG-2021
        \\ \hline\hline
      \end{tabular}
    } %%% \resizebox
    \caption{Exclusive $\Vcb$ in units of $10^{-3}$.}
    \label{tab:ex-Vcb}
  \end{subtable}
  \begin{subtable}{1.0\linewidth}
    \renewcommand{\arraystretch}{1.2}
    \center
%    \vspace*{-7mm}
    \resizebox{1.0\textwidth}{!}{
      \begin{tabular}{ @{\qquad} l @{\qquad\qquad\qquad} l @{\qquad\qquad\qquad} l @{\qquad\qquad\qquad} l @{\qquad} }
        \hline\hline
        channel        & value         & ref  & source \\ \hline
        kinetic scheme & $42.19(78)$   & \cite{ HFLAV:2019otj, Amhis:2016xyh} & HFLAV-2021
        \\
        kinetic scheme & \red{$42.00(64)$} & \cite{ Aoki:2021kgd} p192 & FLAG-2021 
        \\ \hline
        1S scheme      & $41.98(45)$   & \cite{ HFLAV:2019otj, Amhis:2016xyh} & HFLAV-2021 
        \\ \hline\hline
      \end{tabular}
    } %%% \resizebox
    \caption{Inclusive $\Vcb$ in units of $10^{-3}$.}
    \label{tab:in-Vcb}
  \end{subtable}
  \caption{ Results for (\subref{tab:ex-Vcb}) exclusive $\Vcb$ and
    (\subref{tab:in-Vcb}) inclusive $\Vcb$. The same notation as in
    Table \ref{tab:xi0-sum} is used. }
  \label{tab:Vcb}
\end{table}

\section{Input parameters: Wolfenstein parameters}
In Table \ref{tab:input-WP-eta}\,(\subref{tab:WP}), we present the
Wolfenstein parameters on the market.
As explained in Ref.~\cite{ Bailey:2018feb, Bailey:2015frw}, we use
the results of angle-only-fit (AOF) in Table
\ref{tab:input-WP-eta}\,(\subref{tab:WP}) in order to avoid unwanted
correlation between $(\epsK, \Vcb)$, and $(\bar\rho, \bar\eta)$.
We determine $\lambda$ from $\Vus$ which is obtained from the $K_{\ell
  2}$ and $K_{\ell 3}$ decays using lattice QCD inputs for form
factors and decay constants.
We determine the $A$ parameter from $\Vcb$.

\begin{table}[h!]
  \begin{subtable}{0.73\linewidth}
    \renewcommand{\arraystretch}{1.2}
%    \vspace*{-4mm}
    \resizebox{1.0\linewidth}{!}{
      \begin{tabular}{ @{\qquad} c @{\qquad} | l l | l l | l l }
        \hline\hline
        WP
        & \multicolumn{2}{c|}{CKMfitter}
        & \multicolumn{2}{c|}{UTfit}
        & \multicolumn{2}{c}{AOF}
        \\ \hline
        $\lambda$
        & $0.22475(25)$        & \cite{Charles:2004jd}
        & $0.22500(100)$       & \cite{Bona:2006ah}
        & \wlee{ $0.2249(5)$ } & \cite{Aoki:2021kgd} p80
        \\ \hline
        $\bar{\rho}$
        & $0.1577(96)$ & \cite{Charles:2004jd}
        & $0.148(13)$  & \cite{Bona:2006ah}
        & $0.146(22)$  & \cite{Martinelli:2017}
        \\ \hline
        $\bar{\eta}$
        & $0.3493(95)$ & \cite{Charles:2004jd}
        & $0.348(10)$  & \cite{Bona:2006ah}
        & $0.333(16)$  & \cite{Martinelli:2017}
        \\ \hline\hline
      \end{tabular}
    } % resizebox
    \caption{Wolfenstein parameters}
    \label{tab:WP}
  \end{subtable} %%% \subtable
  \hfill
  \begin{subtable}{0.26\linewidth}
    \renewcommand{\arraystretch}{1.2}
%    \vspace*{-4mm}
    \resizebox{1.0\linewidth}{!}{
      \begin{tabular}[b]{ c l c }
        \hline\hline
        Input & Value & Ref.
        \\ \hline
        $\eta_{cc}$ & $1.72(27)$   & \cite{Bailey:2015tba}
        \\
        $\eta_{tt}$ & $0.5765(65)$ & \cite{Buras2008:PhysRevD.78.033005}
        \\
        $\eta_{ct}$ & $0.496(47)$  & \cite{Brod2010:prd.82.094026}
        \\ \hline\hline
      \end{tabular}
    } % resizebox
    \caption{$\eta_{ij}$}
    \label{tab:eta}
  \end{subtable} %%% \subtable
  \caption{ (\subref{tab:WP}) Wolfenstein parameters and
    (\subref{tab:eta}) QCD corrections: $\eta_{ij}$ with $i,j = c,t$.}
  \label{tab:input-WP-eta}
\end{table}

\section{Input parameters: $\BK$, $\xi_\text{LD}$, and others}
In FLAG 2021 \cite{ Aoki:2021kgd}, they report lattice QCD results for
$\BK$ with $N_f=2$, $N_f=2+1$, and $N_f = 2+1+1$.
Here, we use the results for $\BK$ with $N_f=2+1$, which is obtained
by taking an average over the four data points from BMW 11, Laiho 11,
RBC-UKQCD 14, and SWME 15 in Table
\ref{tab:input-BK-other}\;(\subref{tab:BK}).

\begin{table}[htbp]
  \begin{subtable}{0.40\linewidth}
    \renewcommand{\arraystretch}{1.45}
%    \vspace*{-5mm}
    \resizebox{1.0\linewidth}{!}{
      \begin{tabular}{ l  c  l }
        \hline\hline
        Collaboration & Ref. & $\BK$  \\ \hline
        SWME 15       & \cite{Jang:2015sla} & $0.735(5)(36)$     \\
        RBC/UKQCD 14  & \cite{Blum:2014tka} & $0.7499(24)(150)$  \\
        Laiho 11      & \cite{Laiho:2011np} & $0.7628(38)(205)$  \\
        BMW 11        & \cite{Durr:2011ap}  & $0.7727(81)(84)$  \\ \hline
        FLAG 2021     & \cite{Aoki:2021kgd} & $0.7625(97)$
        \\ \hline\hline
      \end{tabular}
    } % resizebox
    \caption{$\BK$}
    \label{tab:BK}
  \end{subtable} %%% \subtable
  \hfill
  \begin{subtable}{0.60\linewidth}
    \renewcommand{\arraystretch}{1.2}
%    \vspace*{-5mm}
    \resizebox{1.0\linewidth}{!}{
      \begin{tabular}{ @{\qquad} c @{\qquad} l @{\qquad} l @{\qquad} }
        \hline\hline
        Input & Value & Ref. \\ \hline
        $G_{F}$
        & $1.1663787(6) \times 10^{-5}$ GeV$^{-2}$
        & PDG-21 \cite{Zyla:2020zbs} \\ \hline
        $M_{W}$
        & $80.379(12)$ GeV
        & PDG-21 \cite{Zyla:2020zbs}\\ \hline
        $\theta$
        & $43.52(5)^{\circ}$
        & PDG-21 \cite{Zyla:2020zbs} \\ \hline
        $m_{K^{0}}$
        & $497.611(13)$ MeV
        & PDG-21 \cite{Zyla:2020zbs} \\ \hline
        $\Delta M_{K}$
        & $3.484(6) \times 10^{-12}$ MeV
        & PDG-21 \cite{Zyla:2020zbs} \\ \hline
        $F_K$
        & $155.7(3)$ MeV
        & FLAG-21 \cite{Aoki:2021kgd} \\ \hline\hline
      \end{tabular}
    } % resizebox
    \caption{Other parameters}
    \label{tab:other}
  \end{subtable} %%% \subtable
  \caption{ (\subref{tab:BK}) Results for $\BK$ and
    (\subref{tab:other}) other input parameters.}
  \label{tab:input-BK-other}
\end{table}

The dispersive long distance (LD) effect is defined as
\begin{align}
  \xi_\text{LD} &=  \frac{m^\prime_\text{LD}}{\sqrt{2} \Delta M_K} \,,
  \qquad
  m^\prime_\text{LD}
  = -\Im \left[ \mathcal{P}\sum_{C}
    \frac{\mate{\wbar{K}^0}{H_\text{w}}{C} \mate{C}{H_\text{w}}{K^0}}
         {m_{K^0}-E_{C}}  \right]
  \label{eq:xi-LD}
\end{align}
As explained in Refs.~\cite{ Bailey:2018feb}, there are two
independent methods to estimate $\xi_\text{LD}$: one is the BGI
estimate \cite{ Buras:2010}, and the other is the RBC-UKQCD estimate
\cite{ Christ:2012, Christ:2014qwa}.
The BGI method is to estimate the size of $\xi_\text{LD}$ using
chiral perturbation theory as follows,
\begin{align}
  \xi_\text{LD} &= -0.4(3) \times \frac{\xi_0}{ \sqrt{2} }
  \label{eq:xiLD:bgi}
\end{align}
The RBC-UKQCD method is to estimate the size of $\xi_\text{LD}$
as follows,
\begin{align}
  \xi_\text{LD} &= (0 \pm 1.6)\%.
  \label{eq:xiLD:rbc}
\end{align}
Here, we use both methods to estimate the size of $\xi_\text{LD}$.

In Table \ref{tab:input-WP-eta}\;(\subref{tab:eta}), we present higher
order QCD corrections: $\eta_{ij}$ with $i,j = t,c$.
A new approach using $u-t$ unitarity instead of $c-t$ unitarity
appeared in Ref.~\cite{ Brod:2019rzc}, which is supposed to have a
better convergence with respect to the charm quark mass.
But we have not incorporated this into our analysis yet, which we will
do in near future.

In Table \ref{tab:input-BK-other}\;(\subref{tab:other}), we present other
input parameters needed to evaluate $\epsK$.

\section{Quark mass}
In Table \ref{tab:m_c:m_t}, we present the charm quark mass $m_c(m_c)$
and top quark mass $m_t(m_t)$.
From FLAG 2021 \cite{ Aoki:2021kgd}, we take the results for $m_c
(m_c)$ with $N_f = 2+1$, since there is some inconsistency among the
lattice results of various groups with $N_f = 2+1+1$.
For the top quark mass, we use the PDG 2021 results for the pole mass
$M_t$ to obtain $m_t (m_t)$.
\begin{table}[t!]
  \begin{subtable}{0.46\linewidth}
    \vspace*{-5mm}
    \renewcommand{\arraystretch}{1.2}
    \resizebox{0.99\linewidth}{!}{
      \begin{tabular}{ l l l l }
        \hline\hline
        {\small Collaboration} & $N_f$ & $m_c(m_c)$ & Ref.
        \\ \hline
        FLAG 2021       & $2+1$   & \red{$1.275(5)$}  & \cite{Aoki:2021kgd}
        \\
        FLAG 2021       & $2+1+1$ & $1.278(13)$ & \cite{Aoki:2021kgd}
        \\ \hline\hline
      \end{tabular}
    } % resizebox
    \caption{$m_c(m_c)$ [GeV]}
    \label{tab:m_c}
  \end{subtable} %%% \subtable
  \hfill
  \begin{subtable}{0.52\linewidth}
    \renewcommand{\arraystretch}{1.2}
    \vspace*{-5mm}
    \resizebox{0.99\linewidth}{!}{
      \begin{tabular}{ l l l l }
        \hline\hline
        {\small Collaboration} & $M_t$ & $m_t(m_t)$ & Ref.
        \\ \hline
        PDG 2019 & 172.9(4)   & 163.08(38)(17)         & \cite{Tanabashi:2018oca} \\
        PDG 2021 & 172.76(30) & \red{ 162.96(28)(17) } & \cite{Zyla:2020zbs}
       \\ \hline\hline
      \end{tabular}
    } % resizebox
    \caption{ $m_t(m_t)$ [GeV] }
    \label{tab:m_t}
  \end{subtable} %%% \subtable
  \caption{  Results for (\subref{tab:m_c}) charm quark mass and
    (\subref{tab:m_t}) top quark mass. }
  \label{tab:m_c:m_t}
\end{table}

\begin{table}[h!]
  \begin{subfigure}{0.55\linewidth}
    \vspace{-3mm}
    \includegraphics[width=1.0\linewidth]{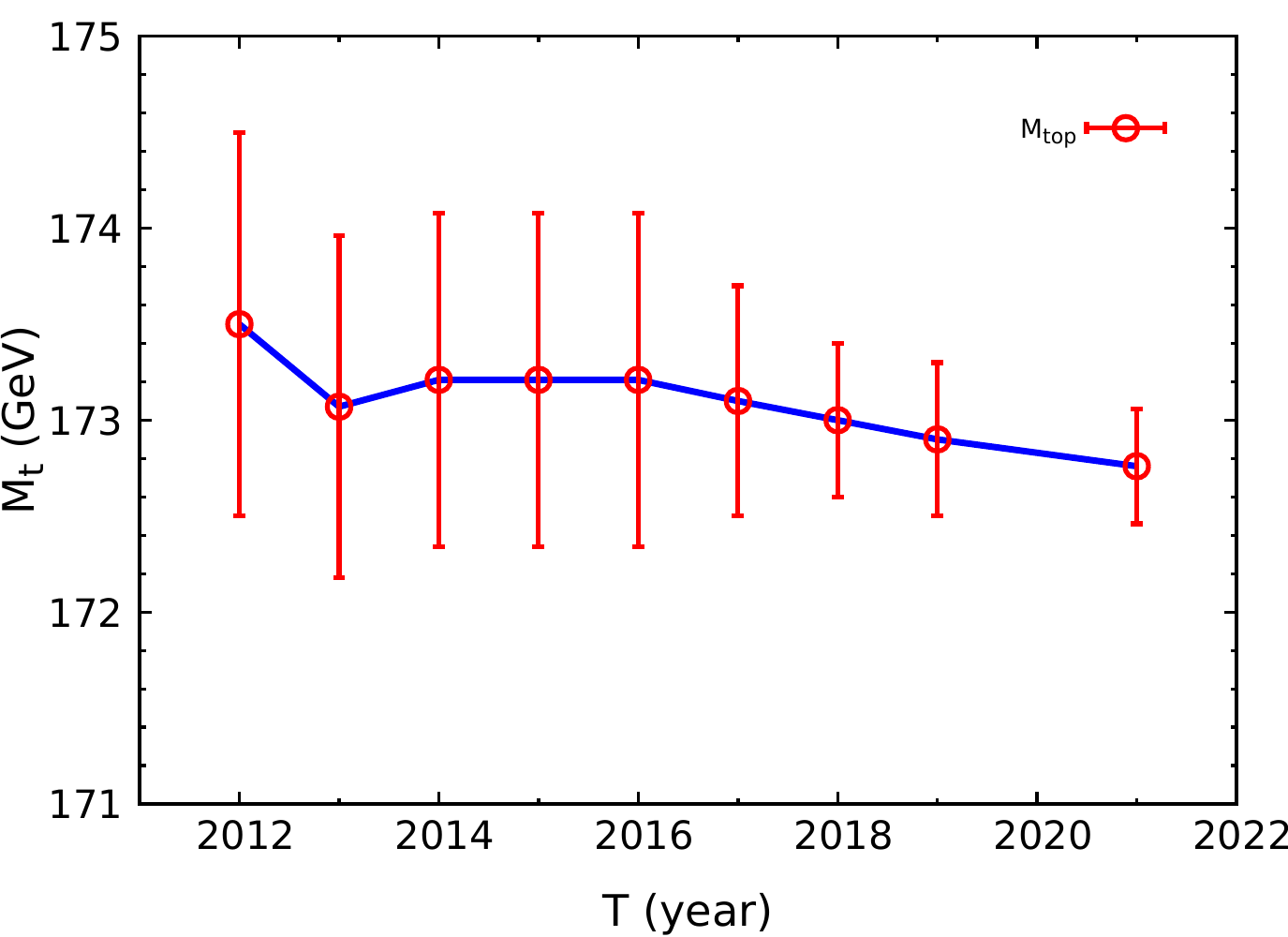}
    \caption{History of $M_t$ (top quark pole mass).}
    \label{fig:M_t}
  \end{subfigure}
  \hfill
  \begin{subtable}{0.45\linewidth}
    \vspace{-2mm}
    \renewcommand{\arraystretch}{1.17}
    \resizebox{0.99\linewidth}{!}{
    \begin{tabular}{@{\qquad} l @{\qquad} l @{\qquad} l }
      \hline\hline
      source & error (\%) & memo 
      \\ \hline
      $\Vcb$          & 44.7             & Exclusive \\
      $\bar\eta$      & 21.4             & AOF \\
      $\eta_{ct}$     & 18.1             & $c-t$ Box \\        
      $\eta_{cc}$     & \phantom{0}7.7   & $c-c$ Box \\        
      $\bar\rho$      & \phantom{0}3.2   & $c-c$ Box \\        
      $\xi_\text{LD}$ & \phantom{0}1.9   & RBC-UKQCD \\        
      $\BK$           & \phantom{0}1.5   & FLAG \\
      $\;\vdots$      & $\;\;\;\;\vdots$ & $\;\;\vdots$
      \\ \hline\hline
    \end{tabular}
    } % resizebox
    \vspace*{4mm}
    \caption{Error budget for $|\epsK|^\text{SM}$}
    \label{tab:err-bud}
  \end{subtable}
  \caption{(\subref{fig:M_t}) $M_t$ history (\subref{tab:err-bud})
    error budget. }
  \label{tab:M_t+err_bud}
\end{table}

In Table \ref{tab:M_t+err_bud} Fig.~(\subref{fig:M_t}), we present the time
evolution of top pole mass $M_t$.
Here we find that the average value drifts downward a little bit and
the error shrinks fast as time goes on, since LHC has been
accumulating high statistics on $M_t$.
The data for 2020 is dropped out intentionally to reflect on the
absence of Lattice 2020 due to COVID-19, even though it is available.

\section{Results for $\epsK$}
\begin{figure}[t!]
  \begin{subfigure}{0.47\linewidth}
    \vspace*{-5mm}
    \includegraphics[width=1.0\linewidth]
       {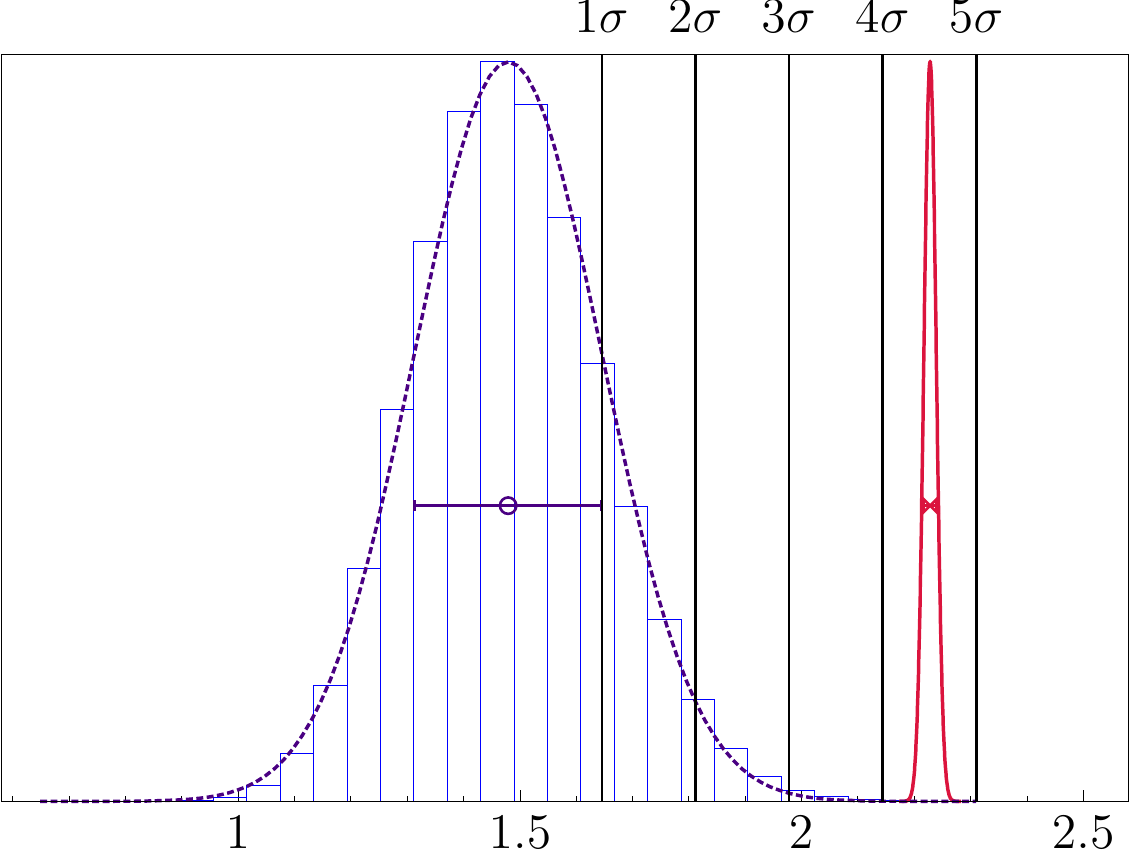}
    \caption{Exclusive $\Vcb$ (FNAL/MILC 2021, BGL)}
    \label{fig:epsK-ex:rbc}
  \end{subfigure}
  \hfill
  \begin{subfigure}{0.47\linewidth}
    \vspace*{-5mm}
    \includegraphics[width=1.0\linewidth]
                    {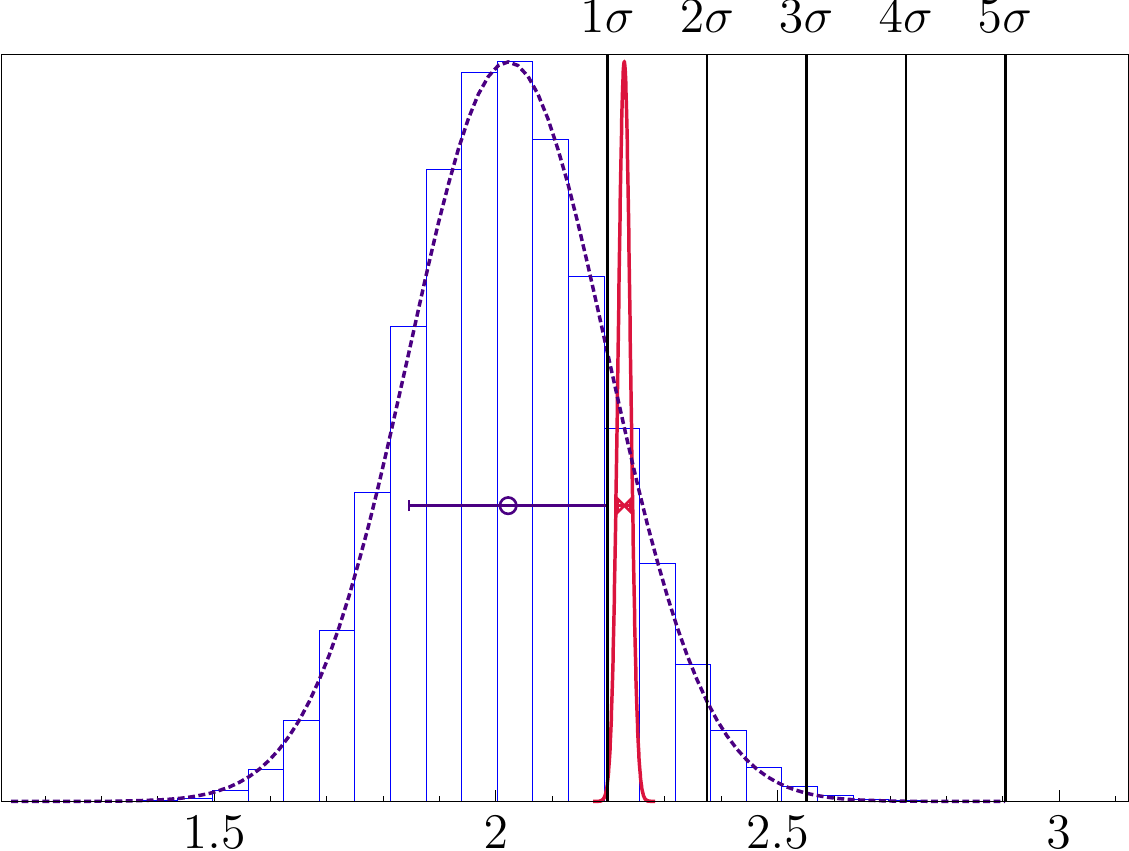}
    \caption{Inclusive $\Vcb$ (HFLAV 2021, 1S scheme)}
    \label{fig:epsK-in:rbc}
  \end{subfigure}
  \caption{$|\epsK|$ with (\subref{fig:epsK-ex:rbc}) exclusive $\Vcb$
    (left) and (\subref{fig:epsK-in:rbc}) inclusive $\Vcb$ (right) in
    units of $1.0\times 10^{-3}$. }
  \label{fig:epsK:cmp:rbc}
\end{figure}

In Fig.~\ref{fig:epsK:cmp:rbc}, we show results for $|\epsK|$
evaluated directly from the standard model (SM) with lattice QCD
inputs given in the previous sections.
In Fig.~\ref{fig:epsK:cmp:rbc}\;(\subref{fig:epsK-ex:rbc}), the blue
curve represents the theoretical evaluation of $|\epsK|$ obtained
using the FLAG-2021 results for $\BK$, AOF for Wolfenstein parameters,
the [FNAL/MILC 2021, BGL] results for exclusive $\Vcb$, results for
$\xi_0$ with the indirect method, and the RBC-UKQCD estimate for
$\xi_\text{LD}$.
The red curve in Fig.~\ref{fig:epsK:cmp:rbc} represents the experimental
results for $|\epsK|$.
In Fig.~\ref{fig:epsK:cmp:rbc}\;(\subref{fig:epsK-in:rbc}), the blue
curve represents the same as in
Fig.~\ref{fig:epsK:cmp:rbc}\;(\subref{fig:epsK-ex:rbc}) except for
using the 1S scheme results for the inclusive $\Vcb$.

Our results for $|\epsK|^\text{SM}$ and $\Delta\epsK$ are summarized
in Table \ref{tab:epsK}.
Here, the superscript ${}^\text{SM}$ represents the theoretical
expectation value of $|\epsK|$ obtained directly from the SM.
The superscript ${}^\text{Exp}$ represents the experimental value
of $|\epsK| = 2.228(11) \times 10^{-3}$.
Results in Table \ref{tab:epsK}\;(\subref{tab:epsK:rbc}) are obtained
using the RBC-UKQCD estimate for $\xi_\text{LD}$, and those in
Table \ref{tab:epsK}\;(\subref{tab:epsK:bgi}) are obtained using
the BGI estimate for $\xi_\text{LD}$.
In Table \ref{tab:epsK}\;(\subref{tab:epsK:rbc}), we find that the
theoretical expectation values of $|\epsK|^\text{SM}$ with lattice QCD
inputs (with exclusive $\Vcb$) has $4.54\sigma \sim 3.68\sigma$ tension
with the experimental value of $|\epsK|^\text{Exp}$, while there is no
tension with inclusive $\Vcb$ (obtained using heavy quark expansion
and QCD sum rules).

\begin{table}[tbhp]
%  \footnotesize
%%%  \renewcommand{\arraystretch}{1.2}
%%%  \renewcommand{\subfigcapskip}{0.55em}
%
  \begin{subtable}{1.0\linewidth}
    \vspace*{-5mm}
    \center
    \renewcommand{\arraystretch}{1.2}
    \resizebox{0.85\linewidth}{!}{
      \begin{tabular}{@{\qquad} l @{\qquad} l @{\qquad} l @{\qquad} l @{\qquad} l @{\qquad} }
        \hline\hline
        $\Vcb$    & method   & reference & $|\epsK|^\text{SM}$ & $\Delta\epsK$
        \\ \hline
        exclusive & CLN      & BELLE 2021 & $1.542 \pm 0.181$  & $3.79\sigma$
        \\
        exclusive & BGL      & BELLE 2021 & $1.528 \pm 0.190$  & $3.68\sigma$
        \\ \hline
        exclusive & CLN      & BABAR 2019 & $1.456 \pm 0.170$  & $4.54\sigma$
        \\
        exclusive & BGL      & BABAR 2019 & $1.451 \pm 0.176$  & $4.42\sigma$
        \\ \hline
        exclusive & CLN      & HFLAV 2021 & $1.577 \pm 0.155$  & $4.21\sigma$
        \\
        exclusive & BGL      & FNAL/MILC 2021 & $1.479 \pm 0.166$ & $4.50\sigma$
        \\ \hline\hline
        inclusive & kinetic  & FLAG 2021  & $2.027 \pm 0.195$ & $1.03\sigma$
        \\
        inclusive & 1S       & HFLAV 2021 & $2.022 \pm 0.176$ & $1.17\sigma$
        \\ \hline\hline
      \end{tabular}
    } % resizebox
    \caption{RBC-UKQCD estimate for $\xi_\text{LD}$}
    \label{tab:epsK:rbc}
  \end{subtable} %%% \subtable
  \begin{subtable}{1.0\linewidth}
    \vspace*{3mm}
    \center
    \renewcommand{\arraystretch}{1.2}
    \resizebox{0.85\linewidth}{!}{
      \begin{tabular}{@{\qquad} l @{\qquad} l @{\qquad} l @{\qquad} l @{\qquad} l @{\qquad} }
        \hline\hline
        $\Vcb$    & method   & reference  & $|\epsK|^\text{SM}$ & $\Delta\epsK$
        \\ \hline
        exclusive & CLN & HFLAV 2021     & $1.625 \pm 0.157$ & $3.85\sigma$
        \\
        exclusive & BGL & FNAL/MILC 2021 & $1.527 \pm 0.169$ & $4.15\sigma$
        \\ \hline\hline
      \end{tabular}
    } % resizebox
    \caption{BGI estimate for $\xi_\text{LD}$}
    \label{tab:epsK:bgi}
  \end{subtable} %%% \subtable
  \caption{ $|\epsK|$ in units of $1.0\times 10^{-3}$, and
    $\Delta\epsK = |\epsK|^\text{Exp} - |\epsK|^\text{SM}$.}
  \label{tab:epsK}
\end{table}

\begin{figure}[htbp]
  \begin{subfigure}{0.501\linewidth}
    %    \vspace*{-7mm}
    \includegraphics[width=\linewidth]
                    {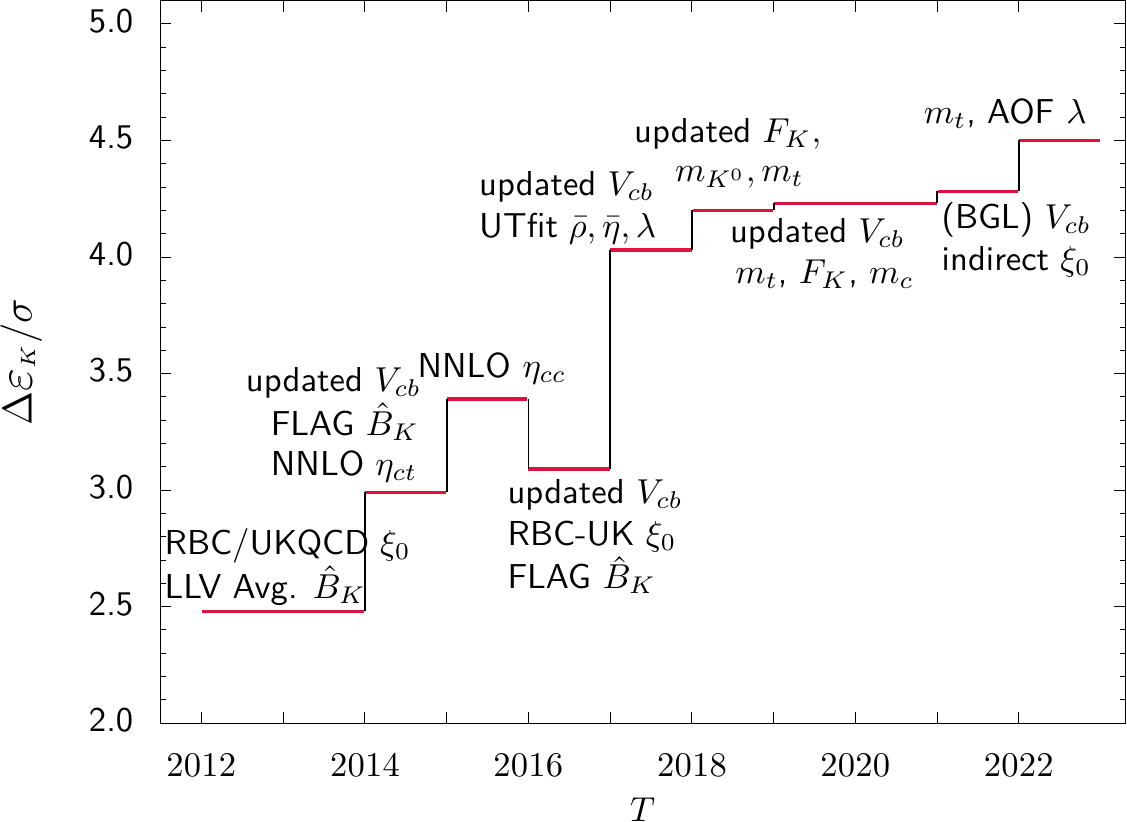}
    \caption{Time evolution of $\Delta \epsK/\sigma$}
    \label{fig:depsK:rbc:his}
  \end{subfigure}
  \hfill
  \begin{subfigure}{0.479\linewidth}
    %    \vspace*{-7mm}
    \includegraphics[width=\linewidth]
                    {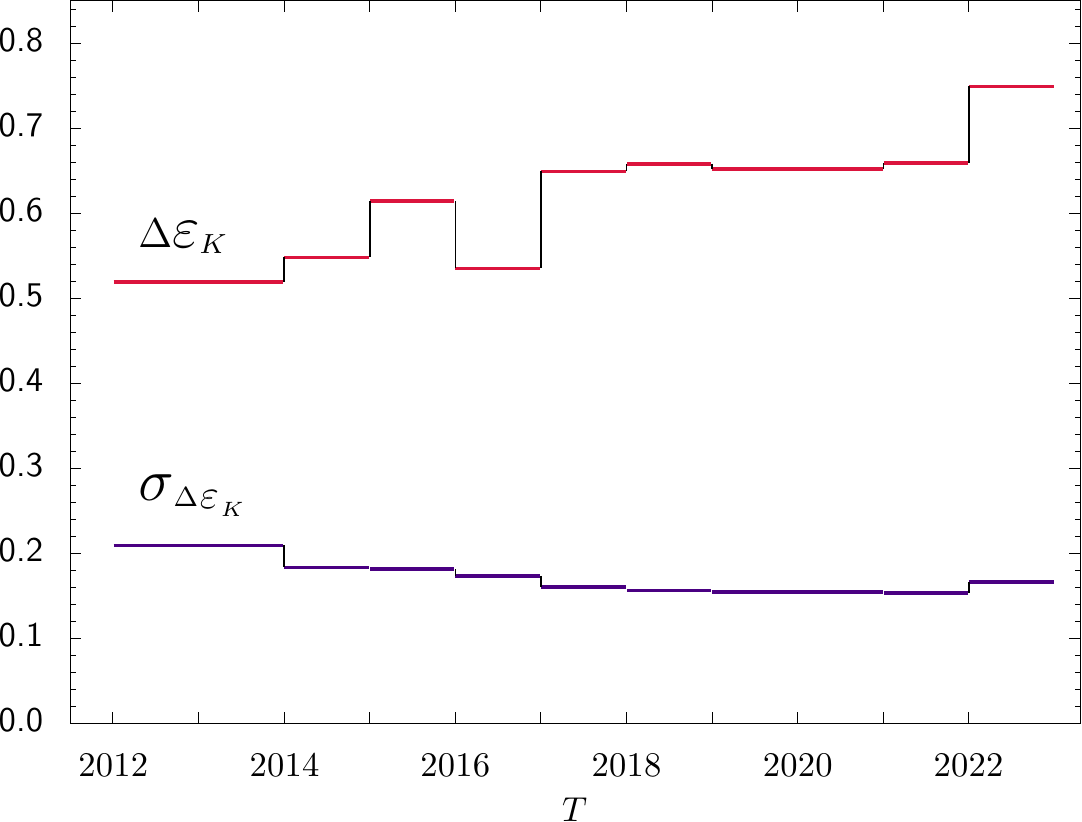}
    \caption{Time evolution of the average and error of $\Delta\epsK$}
    \label{fig:depsK+sigma:rbc:his}
  \end{subfigure}
  \caption{ Time history of (\subref{fig:depsK:rbc:his})
    $\Delta\epsK/\sigma$, and (\subref{fig:depsK+sigma:rbc:his})
    $\Delta\epsK$ and $\sigma_{\Delta\epsK}$. }
  \label{fig:depsK:sum:rbc:his}
\end{figure}

In Fig.~\ref{fig:depsK:sum:rbc:his}\;(\subref{fig:depsK:rbc:his}), we
show the time evolution of $\Delta\epsK$ starting from 2012 to 2022.
In 2012, $\Delta\epsK$ was $2.5\sigma$, but now it is $4.5\sigma$
with exclusive $\Vcb$ (FNAL/MILC 2021, BGL).
In Fig.~\ref{fig:depsK:sum:rbc:his}\;(\subref{fig:depsK+sigma:rbc:his}),
we show the time evolution of the average $\Delta\epsK$ and the error
$\sigma_{\Delta\epsK}$ during the period of 2012--2022.

At present, we find that the largest error ($\approx 45\%$) in
$|\epsK|^\text{SM}$ comes from $\Vcb$.\footnote{Refer to Table
\ref{tab:M_t+err_bud} (\subref{tab:err-bud}) for more details.}
Hence, it is essential to reduce the error in $\Vcb$ significantly.
To achieve this goal, there is an on-going project to extract
exclusive $\Vcb$ using the Oktay-Kronfeld (OK) action for the heavy
quarks to calculate the form factors for $\BtoDstp$ decays \cite{
  Ben:2019, Seungyeob:2019, Bhattacharya:2018ibo, Bailey:2017xjk,
  Bailey:2017zgt, Bailey:2020uon}.

A large portion of interesting results for $|\epsK|^\text{SM}$ and
$\Delta\epsK$ could not be presented in Table \ref{tab:epsK} and in
Fig.~\ref{fig:depsK:sum:rbc:his} due to lack of space: for example,
results for $|\epsK|^\text{SM}$ obtained using exclusive $\Vcb$ (FLAG
2021), results for $|\epsK|^\text{SM}$ obtained using $\xi_0$
determined by the direct method, and so on.
We plan to report them collectively in Ref.~\cite{ wlee:2022epsK}.

%
%\red{EDIT by wlee}
%

\acknowledgments
We thank Jon Bailey, Stephen Sharpe, and Rajan Gupta for helpful
discussion.
The research of W.~Lee is supported by the Mid-Career Research
Program (Grant No.~NRF-2019R1A2C2085685) of the NRF grant funded by
the Korean government (MOE).
This work was supported by Seoul National University Research Grant in
2019.
W.~Lee would like to acknowledge the support from the KISTI
supercomputing center through the strategic support program for the
supercomputing application research (No.~KSC-2018-CHA-0043,
KSC-2020-CHA-0001).
Computations were carried out in part on the DAVID cluster at Seoul
National University.

%------------
% references
%------------
\bibliography{refs}

%----------
% all done
%----------

\end{document}